\documentclass[a4paper,11pt]{article}
\pdfoutput=1 
\usepackage{jheppub} 
\usepackage[T1]{fontenc} 
\usepackage{graphicx,textcomp,calrsfs,calrsfs}

\newcommand{\etal}{{et al.}}
\newcommand{\ie}{{i.e.}, }

\newcommand{\msbar}{\overline{\mbox{\small MS}}}
\newcommand{\be}{\begin{equation}}
\newcommand{\ee}{\end{equation}}
\newlength{\myscale}
\title{\boldmath Weak Mixing Angle in the Thomson Limit}

\author{Jens Erler}
\author{and Rodolfo Ferro-Hern\'andez}

\affiliation{Departamento de F\'isica Te\'orica, Instituto de F\'isica, Universidad Nacional Aut\'onoma de M\'exico, \\
Circuito de la Investigaci\'on Cient\'ifica s/n, Ciudad Universitaria, Coyoac\'an, CDMX, M\'exico}

\emailAdd{erler@fisica.unam.mx}
\emailAdd{ferrohr@estudiantes.fisica.unam.mx}

\abstract{We present a calculation of the weak mixing angle in the $\msbar$ renormalization scheme 
which is relevant for experiments performed at very low energies or momentum transfers.
We include higher orders in the perturbative QCD expansion, 
as well as updated phenomenological and theoretical input, 
and obtain the result $\sin^{2}\hat{\theta}_W(0) = 0.23868(5)(2)$
for the reference values $\hat\alpha_s(M_Z) = 0.1182$ and $\hat m_c(\hat m_c) = 1.272$~GeV.
The first quoted error is from the current Standard Model evaluation of the mixing angle at the $Z$~boson mass scale.
The second error represents the theoretical and parametric uncertainties induced by the evolution to the Thomson limit
and is discussed in detail.}

\begin{document} 
\maketitle
\flushbottom

\section{\label{sec:level1}Introduction}
The electroweak sector of the Standard Model is based on the gauge symmetry group
$SU\left(2\right)_{L} \times U\left(1\right)_{Y}$. 
The weak mixing angle, $\theta_{W}$, is a parameter that describes the mixing of the gauge bosons 
related to the $U\left(1\right)_{Y}$ and the third component of $SU\left(2\right)_{L}$ 
to give rise to the mass eigenstates of the photon and the $Z$ boson. 
In terms of the couplings $g$ of $SU\left(2\right)_{L}$ and $g^\prime$ of $U(1)_Y$ one has
\begin{equation}
\hat{s}^2\equiv\sin^2\hat{\theta}_{W} = \frac{{g^\prime}^2}{g^2 + {g^\prime}^2}\ .
\end{equation}
Since it is given explicitly by gauge couplings, $\sin^2\hat{\theta}_{W}$ depends on the energy scale 
and is governed by a renormalization group equation (RGE)~\cite{Czarnecki:2000ic,Erler:2004in}. 

One of the tests of the Standard Model is to evolve the weak mixing angle from high to low energies 
and compare it with experimental extractions at lower squared momentum transfers, $Q^2$.
For example, the Qweak experiment~\cite{Androic:2013rhu} at Jefferson Laboratory (JLab) has measured 
the weak charge of the proton, $Q_W(p) \sim 1 - 4 \sin^2\theta_W$, in polarized 
electron scattering from a fixed liquid hydrogen target at $Q^2 \approx 0.026$~GeV$^2$.
The same observable, but at an even lower $Q^2 \approx 0.0045$~GeV$^2$, will also be targeted 
by the P2 experiment~\cite{Becker:2018ggl} at the MESA facility 
which is currently under construction at the University of Mainz in Germany. 
In a very similar setup, the MOLLER Collaboration~\cite{Benesch:2014bas} at JLab will build and 
improve on the completed E158 experiment~\cite{Anthony:2005pm} at SLAC 
(that occurred at almost the same $Q^2$ as Qweak) 
and measure the analogous weak charge of the electron, $Q_W(e)$, 
in polarized M\o ller scattering at $Q^2 \approx 0.0056$~GeV$^2$.
The PVDIS Collaboration~\cite{Wang:2014bba} at the 6~GeV CEBAF complex at JLab scattered polarized 
electrons deep-inelastically from deuterium, 
and the SoLID Collaboration~\cite{Souder:2016xcn} will increase the PVDIS precision in the future by 
benefiting from the energy upgraded CEBAF and a correspondingly higher and broader $Q^2$ range.
Other approaches include neutrino and anti-neutrino deep inelastic scattering~\cite{Zeller:2001hh},
$\bar\nu$-$e$ scattering near nuclear reactors~\cite{Canas:2016vxp}, 
and parity violation in atoms~\cite{Wood:1997zq} and ions~\cite{Willmann:2012mxa}.
For more details, see the recent reviews on 
low energy measurements of the weak mixing angle~\cite{Kumar:2013yoa}, 
on the weak neutral current~\cite{Erler:2013xha}, 
and on weak polarized electron scattering~\cite{Erler:2014fqa}.

Since QCD at low energies does not allow for reliable perturbative calculations, the theoretical uncertainty 
of the RGE running from the $Z$-pole to low energies arises dominantly from the hadronic region. 
A phenomenological approach to address this region was developed in Ref.~\cite{Erler:2004in}. 
Working in the $\msbar$ scheme\footnote{Quantities defined in the $\msbar$ scheme will be denoted
by a caret.}, the main idea was to relate the case of the weak mixing angle to that 
of the electromagnetic coupling, $\hat\alpha$, as far as possible, and then to consider both maximal and 
minimal $SU(3)$ flavor symmetry breaking to constrain the flavor separation of the three light quarks $(u,d,s)$.
In the present work, we extend the analysis to the next order in the strong coupling constant, $\hat\alpha_s$, 
and introduce a number of new elements. 
We employ the most recent values and uncertainties of the input parameters, such as $\hat\alpha_s$ and
the heavy quark masses.
The hadronic vacuum polarization contribution to the RGE running of $\hat\alpha$ is obtained 
dispersively from $e^+ e^-$ annihilation data for hadronic final states, 
which are supplemented by isospin rotated $\tau$ decay spectral functions 
corrected for isospin breaking effects~\cite{Davier:2017zfy,Jegerlehner:2011ti}. 
We tie experimental data~\cite{Davier:2017zfy} and lattice gauge theory 
calculations~\cite{Blum:2016xpd,Chakraborty:2014mwa} together to obtain the individual 
contributions of strange and first generation quarks. 
This flavor separation at the quark level to high accuracy is consistent with and almost an order of 
magnitude more precise than previous calculations~\cite{Czarnecki:2000ic,Erler:2004in}. 
It is also necessary to constrain OZI-rule~\cite{Okubo:1963fa,Zweig:1964jf,Iizuka:1966fk} violating 
effects, for which we utilize the recent lattice gauge theory calculation of disconnected 
contributions to the anomalous magnetic moment of the muon~\cite{Blum:2015you}.
These refinements allow for significant reduction of the theoretical uncertainty of the RGE evolution. 
As a by-product, our method sheds light on the dual description of quarks and hadrons in the non-perturbative 
regime and may open new ways to extract the strange quark mass from the electro-production of hadrons.

The paper is organized as follows:  
Section~\ref{sec:RGE} presents the RGE of the weak mixing angle up to five loop accuracy, 
and the matching conditions for $\hat{\alpha}$ and $\sin^{2}\theta_{W}$. 
In Section~\ref{sec:MSbar} we perform the conversion of the hadronic vacuum polarization contribution 
to the running of $\hat\alpha$ from the on-shell scheme, where it is most directly obtained, to the $\msbar$ scheme
(Appendix~\ref{appendixB} contains a brief discussion of various calculations of $\alpha(M_Z)$).
Section~\ref{sec:singlet} describes the calculation of the singlet contribution to the weak mixing angle,  
with some details given in Appendix~\ref{appendixA}.
In Section~\ref{sec:flavor} the flavor separation (contributions of light and strange quarks) is addressed 
and threshold masses are calculated. 
In Section~\ref{sec:uncertainties} theoretical uncertainties are discussed in detail,
and Section~\ref{sec:conclusions} offers our final results and conclusions.

\section{\label{sec:RGE}Renormalization group evolution}
In an approximation in which all fermions are either massless and active or infinitely heavy and decoupled, 
the RGE for the electromagnetic coupling in the $\msbar$ scheme~\cite{Erler:1998sy}, $\hat{\alpha}$, can be 
written in the form~\cite{Erler:2004in},
\begin{equation}
\mu^2 \frac{d\hat\alpha}{d\mu^2} = \frac{\hat\alpha^2}{\pi} \left[\frac{1}{24} \sum_i K_i \gamma_i Q_i^2 + 
\sigma\left(\sum_q Q_q \right)^2 \right],
\label{eq:fine structure}
\end{equation}
where the sum is over all active particles in the relevant energy range.
The $Q_i$ are the electric charges, while the $\gamma_i$ are constants 
depending on the field type and shown in Table~\ref{tab:Numerical-constants--1}.
\begin{table}[t]
\centering
\begin{tabular}{|lrclr|}
\hline 
boson & $\gamma_{i}$ & \phantom{OOOOO} & fermion & $\gamma_{i}$\tabularnewline
\hline 
real scalar & $1$ &  & chiral fermion & $4$\tabularnewline
complex scalar & $2$ &  & Majorana fermion & $4$\tabularnewline
massless gauge boson & $-22$ &  & Dirac fermion & $8$\tabularnewline
\hline 
\end{tabular}
\caption{RGE contributions of different particle types, 
where the minus sign is indicative for the asymptotic freedom in non-Abelian gauge theories.}
\label{tab:Numerical-constants--1}
\end{table}
The $K_i$ and $\sigma$ contain higher-order corrections and are given by~\cite{Baikov:2012zm},
\begin{eqnarray}
\nonumber
K_i &=& N_i^c \left\{ 1 + \frac{3}{4} Q_i^2 \frac{\hat\alpha}{\pi} + \frac{\hat\alpha_s}{\pi} + 
\frac{\hat\alpha_s^2}{\pi^2} \left[ \frac{125}{48} - \frac{11}{72} n_q \right] \right. \\[12pt]
\nonumber
&+& \frac{\hat\alpha_s^3}{\pi^3} \left[ \frac{10487}{1728} + \frac{55}{18} \zeta_3 - 
n_q \left( \frac{707}{864} + \frac{55}{54} \zeta_3 \right)  - \frac{77}{3888} n_q^2 \right] \\[12pt] 
\nonumber 
&+& \frac{\hat\alpha_s^4}{4\pi^4} \left[ \frac{2665349}{41472} + \frac{182335}{864}\zeta_3 - 
\frac{605}{16}\zeta_4 - \frac{31375}{288}\zeta_5 \right. \\[12pt]
\nonumber
&-& n_q \left( \frac{11785}{648} + \frac{58625}{864}\zeta_3 - \frac{715}{48}\zeta_4 - 
\frac{13325}{432}\zeta_5 \right) \\[12pt]
&-& n_q^2 \left( \frac{4729}{31104} - \frac{3163}{1296}\zeta_3 + \frac{55}{72}\zeta_4 \right)  
+ n_q^3 \left.\left. \left( \frac{107}{15552}+\frac{1}{108}\zeta_{3}\right) \right] \right\},
\label{eq:ki}
\end{eqnarray}
and,
\begin{eqnarray}
\nonumber
\sigma = \frac{\hat\alpha_s^3}{\pi^3} \left[ \frac{55}{216} - \frac{5}{9}\zeta_3 \right]
+ \frac{\hat\alpha_s^4}{\pi^4} \left[ \frac{11065}{3456} - \frac{34775}{3456}\zeta_3 + 
\frac{55}{32}\zeta_4 + \frac{3875}{864}\zeta_5 \right. \\[12pt]
- \left. n_q \left( \frac{275}{1728} - \frac{205}{576}\zeta_3 + \frac{5}{48}\zeta_4 + 
\frac{25}{144}\zeta_5 \right)\right],
\label{eq:sigma}
\end{eqnarray}
with $n_q$ the number of active quarks and $N_i^c = 3$ the color factor for quarks.
For leptons one substitutes $N_i^c =1$ and $\hat\alpha_s = 0$, while $K_i = 1$ for bosons. 

\begin{table}
\centering{}%
\begin{tabular}{|ccccc|}
\hline 
\phantom{O} Energy range \phantom{O} & \phantom{O} $\lambda_{1}$ \phantom{O} & \phantom{O} $\lambda_{2}$ 
\phantom{O} & \phantom{O} $\lambda_{3}$ \phantom{O} & \phantom{O} $\lambda_{4}$ \phantom{O}
\tabularnewline\hline\vspace*{-11pt} & & & & \\ 
$\bar{m}_{t}\leq\mu$ \phantom{OO} & $\frac{9}{20}$ & $\frac{289}{80}$ & $\frac{14}{55}$ & $\frac{9}{20}$
\tabularnewline\vspace*{-10pt} & & & & \\ 
$M_{W}\leq\mu<\bar{m}_{t}$ & $\frac{21}{44}$ & $\frac{625}{176}$ & $\frac{6}{11}$ & $\frac{3}{22}$
\tabularnewline\vspace*{-10pt} & & & & \\ 
$\bar{m}_{b}\leq\mu<M_{W}$ & $\frac{21}{44}$ & $\frac{15}{22}$ & $\frac{51}{440}$ & $\frac{3}{22}$
\tabularnewline\vspace*{-10pt} & & & & \\ 
$m_{\tau}\leq\mu<\bar{m}_{b}$ & $\frac{9}{20}$ & $\frac{3}{5}$ & $\frac{2}{19}$ & $\frac{1}{5}$
\tabularnewline\vspace*{-10pt} & & & & \\ 
$\bar{m}_{c}\leq\mu<m_{\tau}$ & $\frac{9}{20}$ & $\frac{2}{5}$ & $\frac{7}{80}$ & $\frac{1}{5}$
\tabularnewline\vspace*{-10pt} & & & & \\ 
$\bar{m}_{s}\leq\mu<\bar{m}_{c}$ & $\frac{1}{2}$ & $\frac{1}{2}$ & $\frac{5}{36}$ & $0$
\tabularnewline\vspace*{-10pt} & & & & \\ 
$\bar{m}_{d}\leq\mu<\bar{m}_{s}$ & $\frac{9}{20}$ & $\frac{2}{5}$ & $\frac{13}{110}$ & $\frac{1}{20}$
\tabularnewline\vspace*{-10pt} & & & & \\ 
$\bar{m}_{u}\leq\mu<\bar{m}_{d}$ & $\frac{3}{8}$ & $\frac{1}{4}$ & $\frac{3}{40}$ & $0$
\tabularnewline\vspace*{-10pt} & & & & \\ 
$m_{\mu}\leq\mu<\bar{m}_{u}$ & $\frac{1}{4}$ & $0$ & $0$ & $0$
\tabularnewline\vspace*{-10pt} & & & & \\ 
$m_{e}\leq\mu<m_{\mu}$ & $\frac{1}{4}$ & $0$ & $0$ & $0$\vspace*{-11pt} \\ & & & &
\tabularnewline
\hline 
\end{tabular}
\protect\caption{Coefficients entering the higher order RGE for the weak mixing angle.}
\label{tab:Numerical-constants--2}
\end{table}

\begin{figure}[t]
\centering
\includegraphics[scale=0.25]{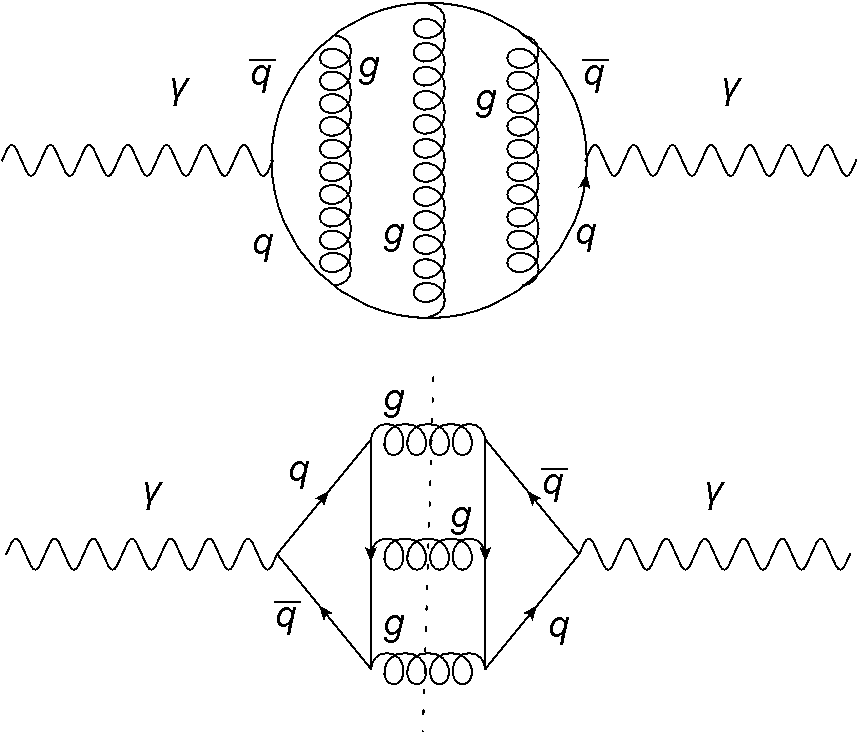}\vspace{-1pt}
\caption{Examples of a connected (top) and a disconnected (bottom) Feynman diagram.}
\label{fig:discdiagram}
\end{figure}

We can relate the RGE of $\hat{\alpha}$ to that of $\sin^2\hat\theta_W$ since both, the $\gamma Z$ mixing 
tensor $\hat\Pi_{\gamma Z}$ and the photon vacuum polarization function $\hat\Pi_{\gamma\gamma}$ are pure 
vector-current correlators. Including higher order corrections, the RGE for the $Z$ boson vector coupling to fermion $f$,
$\hat v_f = T_f - 2 Q_f \sin^2\hat\theta_W$,
where $T_f$ is the third component of weak isospin of fermion $f$, is then
\begin{equation}
\mu^{2}\frac{d\hat{v}_{f}}{d\mu^{2}} = \frac{\hat\alpha Q_f}{24\pi}
\left[ \sum_i K_i \gamma_i \hat{v}_i Q_i + 12 \sigma \left( \sum_q Q_q \right) 
\left( \sum_q \hat{v}_q \right)\right].
\label{eq:weakRGE1}
\end{equation}
Eqs.~(\ref{eq:fine structure}) and (\ref{eq:weakRGE1}) can be used~\cite{Erler:2004in} to obtain
\begin{eqnarray} 
\nonumber
\hat s^2(\mu) = \hat s^2(\mu_0) \frac{\hat\alpha(\mu)}{\hat\alpha(\mu_0)} + \lambda_1 
\left[ 1 - \frac{\hat\alpha(\mu)}{\hat\alpha(\mu_0)} \right] + \\[12pt]
\frac{\hat\alpha(\mu)}{\pi} \left[ \frac{\lambda_2}{3} \ln\frac{\mu^2}{\mu_0^2} + \frac{3\lambda_3}{4} 
\ln \frac{\hat\alpha(\mu)}{\hat\alpha(\mu_0)} + \tilde\sigma(\mu_{0}) - \tilde\sigma(\mu) \right],
\label{eq:MASTEREQUATION}
\end{eqnarray}
where the $\lambda_{i}$ are known~\cite{Erler:2004in} constants given in 
Table~\ref{tab:Numerical-constants--2} and the explicit $K_i$ dependence has disappeared. 
The $\tilde\sigma$ terms,
\begin{equation}
\tilde\sigma(\mu) = \frac{\lambda_4}{33 - 2 n_q}\, \frac{5}{36}
\left[ (11 - 24\zeta_3)\, \frac{\hat\alpha_s^2(\mu)}{\pi^2} + b\, \frac{\hat\alpha_s^3(\mu)}{\pi^3} \right],
\label{eq:OZIterm}
\end{equation}
with,
\begin{eqnarray}
\nonumber
b &\equiv& \frac{2213}{24} - \frac{6955}{24}\zeta_3 + \frac{99}{2}\zeta_4 + \frac{775}{6}\zeta_5 - 
n_q \left( \frac{55}{12} - \frac{41}{4}\zeta_3 + 3\zeta_4 + 5\zeta_5 \right) \\[12pt]
&-& \frac{(153 - 19 n_q)(11 - 24\zeta_3)}{99 - 6 n_q}\ ,
\end{eqnarray} 
represent the singlet contributions to the RGE evolution of the weak mixing angle at four and five loop order. 
These terms arise from quark-antiquark annihilation (disconnected) diagrams (see Figure~\ref{fig:discdiagram}) 
and are suppressed in perturbative QCD (PQCD). 
In the non-perturbative domain these give rise to so-called 
OZI-rule~\cite{Okubo:1963fa,Zweig:1964jf,Iizuka:1966fk} violations. 

Eq.~(\ref{eq:MASTEREQUATION}) together with the solution of the four-loop QCD $\beta$-function~\cite{vanRitbergen:1997va,Czakon:2004bu} represents a complete solution, as long as all matching scales 
$\mu$ at which an active particle decouples are known, because there the $\lambda_i$ change their values.
The matching scales of all bosons~\cite{Hall:1980kf}, charged leptons, and heavy ($t$, $b$, and $c$) 
quarks~\cite{Chetyrkin:1997un,Chetyrkin:2006xg,Kniehl:2006bf} can be calculated as what we call threshold 
masses $\bar m_q$, where the QCD corrections to the matching relations vanish by definition.

\subsection{Matching conditions} 
At each particle threshold the RGE coefficients need to be modified to reflect the particle content of the 
associated effective field theory (EFT), and in the $\msbar$ scheme it is also convenient to change the 
definitions of $\hat\alpha$ and $\hat{s}$ to correspond to this same EFT.
This is analogous to the usual treatment of $\hat\alpha_s$ and leads to very small matching discontinuities in 
the RGE running of the couplings. 

Denoting the electromagnetic coupling with and without the fermion near the threshold by $\hat\alpha(m_f)^+$ 
and $\hat\alpha(m_f)^-$, respectively\footnote{We assume $m_{f}$ is an $\msbar$ mass with respect to QCD, 
but a pole mass for both leptons and quarks with respect to QED.}, 
the matching condition for $\hat{\alpha}$ reads~\cite{Chetyrkin:1997un,Chetyrkin:2006xg,Kniehl:2006bf}, 
\begin{eqnarray}
\frac{\pi}{\hat\alpha(m_f)^+} &=& \frac{\pi}{\hat\alpha(m_f)^-} -
\frac{15}{16} N_f^c \frac{\hat\alpha(m_f)}{\pi} Q_f^4 \nonumber \\[12pt]
&-& \frac{N_f^c - 1}{2} \left[ \frac{13}{12} \frac{\hat\alpha_s^{+}}{\pi} + \left( \frac{655}{144}\zeta_3 - 
\frac{3847}{864} + \frac{361}{1296} n_q \right) \frac{\hat\alpha_s^{+2}}{\pi^2} \right.\nonumber \\[12pt]
&+& \left.\left( - 0.55739 - 0.92807\, n_q + 0.01928\, n_q^2 \right)
\frac{\hat\alpha_s^{+3}}{\pi^3} \right] Q_f^2 \nonumber \\[12pt]
&-& \frac{N_f^c - 1}{2} \left[ \frac{295}{1296} \frac{\hat\alpha_s^{+2}}{\pi^2} + 
({\mathcal K}_1 + {\mathcal K}_2 n_q) \frac{\hat\alpha_s^{+3}}{\pi^3} \right] \sum_{\ell} Q_\ell^2\ . 
\label{eq:PIo}
\end{eqnarray}
The first three lines derive from heavy quark vector-current correlators. 
The last line involves a sum over all quarks $\ell$ with $m_\ell \ll m_q$, and arises from the decoupling of 
the heavy quark $q$ propagating in inner loops of multi-bubble type diagrams in which the outer loop 
(the one coupled to the currents) is occupied by a light quark $\ell$.
The corresponding contribution at order $\hat{\alpha}_{s}^{3}$ is parametrized by the coefficients 
${\mathcal K}_i$ and is unknown at present.
The known $\hat{\alpha}_{s}^2$ term for the charm and bottom quarks, and the $\hat{\alpha}_{s}^3$ terms from 
the charm and bottom quark vector-current correlators amount to about $9 \times 10^{-6}$ and 
$-9 \times 10^{-6}$, respectively.
Taking these as conservative bounds on the unknown higher-order terms and combining them in quadrature 
results in an estimated truncation error of $\pm 1.3\times 10^{-5}$ in $\hat\alpha$.

The matching conditions of $\hat s^2$ and $\hat\alpha$ can also be related~\cite{Erler:2004in}, 
\begin{eqnarray}
\sin^2\hat\theta_W(\hat m_f)^- =
\frac{\hat\alpha(\hat m_f)^-}{\hat\alpha(\hat m_f)^+} \sin^2\hat\theta_W(\hat m_f)^+ 
+ \frac{Q_i T_i}{2 Q_i^2} \left[ 1 - \frac{\hat\alpha(\hat m_f)^-}{\hat\alpha(\hat m_f)^+} \right].
\label{matchs2w}
\end{eqnarray} 
Applying the numerical analysis of the previous paragraph to Eq.~(\ref{matchs2w}), 
we find $2.4 \times 10^{-6}$ and $-1.4 \times 10^{-6}$, respectively,
and we estimate a truncation error related to the matching of about $\pm 3\times 10^{-6}$ in $\hat s^2$.

For completeness we recall that integrating out the $W^\pm$ bosons 
induces the one-loop matching condition~\cite{Erler:2004in,Hall:1980kf},
\begin{equation}
\frac{1}{\hat\alpha^+} = \frac{1}{\hat\alpha^-} + \frac{1}{6\pi}\ .
\end{equation}
For $\hat s^2$ this implies
\begin{equation}
\sin^2\hat\theta_W(M_W)^+ = 1 - \frac{\hat\alpha(M_W)^+}{\hat\alpha(M_W)^-}\cos^2\hat\theta_W(M_W)^- .
\end{equation}
\section{\label{sec:MSbar}Implementation of experimental input}
The perturbative treatment of the previous section cannot be applied at hadronic energy scales 
and experimental input is required.
This is usually taken from $R(s)$, \ie the cross section 
$\sigma(e^+ e^- \to \mbox{ hadrons})$ normalized to $\sigma(e^+ e^- \to \mu^+ \mu^-)$.
Additional information on $R(s)$ is encoded in hadronic $\tau$ decay spectral functions~\cite{Alemany:1997tn}.
The traditional method to implement the $R(s)$ measurements is through a subtracted dispersion integral,
\be 
\Delta\alpha^{(5)}_{\rm had}(M_Z^2) = \frac{\alpha}{3 \pi} \int_{4 m_\pi^2}^\infty
d s \frac{R(s) M_Z^2}{s(M_Z^2 - s) - i \epsilon} \ ,
\label{SDR}
\ee
which gives the hadronic contribution (with the top quark removed) 
to the $Z$ scale value of the electromagnetic coupling in the on-shell scheme.
One supplements the input data with the theoretical (perturbative) prediction for $R(s)$ 
at $s \geq s_0$, with $s_0$ large enough to be able to trust QCD perturbation theory.
A variant~\cite{Eidelman:1998vc} of this approach evaluates Eq.~(\ref{SDR}) in the space-like region, 
$\Delta\alpha^{(5)}_{\rm had.}(-M_Z^2)$, and obtains $\Delta\alpha^{(5)}_{\rm had.}(M_Z^2)$ in a second step. More details about how different groups get the running of alpha are given in Appendix~\ref{appendixB}.

In the $\msbar$ scheme it is more natural to use an unsubstracted dispersion relation~\cite{Erler:1998sy},
\begin{equation}  
\label{eq:delalp3mu0}
\Delta\hat\alpha^{(3)}(\mu_0) = \frac{\alpha}{3\pi} \int_{4 m_\pi^2}^{\mu_0^2}
d s \frac{R(s)}{s - i \epsilon} + 4\pi I^{(3)}, 
\end{equation}
where the superscript indicates that we focus here on the currents produced by the three light quarks
(bosons, leptons, charm and bottom quarks are included following Sec.~\ref{sec:RGE}).
The upper integration limit can in principle be chosen as an arbitrary perturbative scale $\mu_0$, 
but in practice we take $\mu_0^2$ to coincide with the cut-off value $s_0$ used in the traditional method,
since this allows us to recycle results obtained there. 
Indeed~\cite{Erler:1998sy},
\be 
\frac{\alpha}{3\pi} \int\limits_{4 m_\pi^2}^{\mu_0^2} ds 
\left[ \frac{R(s)}{s - i \epsilon} - \frac{R(s) M_Z^2}{s (M_Z^2 - s) - i \epsilon} \right] < 10^{-6} ,
\ee
for $\mu_0 \lesssim 2$~GeV.
Using the results of Ref.~\cite{Davier:2017zfy} including inputs from $\tau$ decays
which we correct for $\gamma$-$\rho$ mixing~\cite{Jegerlehner:2011ti}, we obtain,
\be 
\frac{\alpha}{3\pi} \int\limits_{4 m_\pi^2}^{4~{\rm GeV}^2} ds 
\frac{R(s) M_Z^2}{s (M_Z^2 - s)} = (58.71 \pm 0.45) \times 10^{-4}\ .
\label{eq:noglobalfit}
\ee

We compute the second term in Eq.~(\ref{eq:delalp3mu0}) at the scale $\mu = 2$~GeV 
perturbatively~\cite{Erler:2017dic}, extending the ${\mathcal O}(\hat\alpha_s^2)$ result of 
Ref.~\cite{Erler:1998sy} to ${\mathcal O}(\hat\alpha_s^3)$,
\begin{eqnarray}
4\pi I^{(3)} &=& 2\alpha \int_0^{2\pi} d\theta\, \hat{\Pi}^{(3)}(\mu^2 e^{i\theta}) \nonumber \\[12pt]
&=& \frac{2\alpha}{3\pi} \biggl[ \frac{5}{3} +\left(\frac{55}{12}-4\zeta(3)+2\frac{\hat{m}_{s}^2}{\mu^2}\right)\left( \frac{\hat\alpha_s}{\pi}+\frac{\hat\alpha}{4\pi}\right) \nonumber \\[12pt]
&+& \left(\frac{34525}{864}-\frac{9}{4}\zeta(2)-\frac{715}{18}\zeta(3)+\frac{25}{3}\zeta(5)+\frac{125}{12}\frac{\hat{m}_{s}^2}{\mu^2}+F_2(\hat{m}_c,\hat{m}_b)\right)\, \frac{\hat\alpha_s^2}{\pi^2} \nonumber \\[12pt]
&+& \biggl(\frac{7012579}{13824}-\frac{961}{16}\zeta(2)-\frac{76681 }{144}\zeta (3)+\frac{12515 }{288}\zeta (5)\nonumber \\[12pt]
&&-\frac{665 }{36}\zeta
   (7)+\frac{81}{2}\zeta(2)\zeta(3)+\frac{155
   }{2}\zeta (3)^2+F_3(\hat{m}_c,\hat{m}_b)\biggl)\, \frac{\hat\alpha_s^3}{\pi^3}\biggl] \nonumber \\[12pt]
&=& (24.85 \pm 0.18 - 43\, \Delta\hat\alpha_s) \times 10^{-4},
\label{eq:I3}
\end{eqnarray}
where the $F_i(\hat{m}_c,\hat{m}_b)$ are correction terms from the charm and bottom quarks. 
The explicit analytical expression for ${F_2(\hat{m}_c,\hat{m}_b)\simeq-0.2348}$ is given in 
Ref.~\cite{Erler:1998sy}, 
while that for $F_3(\hat{m}_c,\hat{m}_b)\simeq-0.390$ will appear in ref.~\cite{Erler:2017dic}. 
The numerical evaluation in the last line of Eq.~(\ref{eq:I3}) is for $\hat\alpha_s(M_Z) = 0.1182$,  
$\hat\alpha_s(2~\text{GeV}) = 0.303$ and 
$\hat m_s(2~\mbox{GeV}) = 98 \pm 6$~MeV~\cite{Patrignani:2016xqp}. 
The uncertainty is the size of the 
${\mathcal O}(\hat\alpha_s^3)$ term, and we have defined 
\be 
\Delta\hat\alpha_s \equiv \hat\alpha_s(M_Z) - 0.1182 ,
\label{eq:alphasvar}
\ee
to display the dependence on $\hat\alpha_s$.
Thus, from Eqs.~(\ref{eq:delalp3mu0})--(\ref{eq:I3}) we obtain,
\begin{equation}
\Delta\hat\alpha^{(3)}(2~\text{GeV}) = (83.56 \pm 0.45 \pm 0.18) \times 10^{-4}\ .
\label{eq:deltatwogevms}
\end{equation}

\section{Singlet contribution~\label{sec:singlet}}
We recall that Eq.~(\ref{eq:OZIterm}) exhibits an explicit dependence on $\alpha_s$, which in the non-perturbative domain 
gives rise to the QCD induced OZI-rule~\cite{Okubo:1963fa,Zweig:1964jf,Iizuka:1966fk} violations.
These have to be known independently, since they affect $\hat\alpha$ and $\hat s^2$ differently. 
Thus, in addition to a quark flavor separation, one also needs a singlet piece 
separation, even though the singlet piece is expected to be small. 
To do so, we first relate $\Delta_{\rm disc}\hat{\alpha}$, the disconnected part in $\Delta\hat\alpha^{(3)}(2\mbox{ GeV})$, 
to the one entering the low energy weak mixing angle, $\Delta_{\rm disc}\hat s^2$. 
Non-singlet and singlet contributions are separately gauge-invariant, and 
to gain information on $\Delta_{\rm disc}\hat{\alpha}$, we will adopt a lattice QCD calculation~\cite{Blum:2015you} 
of the disconnected quark line contributions to the anomalous magnetic moment of the muon, $a_\mu$. 

By construction, the $\tilde{\sigma}$ terms in Eq.~(\ref{eq:MASTEREQUATION}) are 
related to the $\sigma$ terms in Eq.~(\ref{eq:fine structure}),
\begin{equation}
\mu^2\frac{d\tilde{\sigma}}{d\mu^2} = - \lambda_4 \sigma.
\label{eq:relationozi}
\end{equation} 
On the other hand, isolating the $\Delta_{\rm disc}\hat{\alpha}$ term in Eq.~(\ref{eq:fine structure}) 
we obtain (working here in lowest order in $\alpha$),
\begin{equation}
\mu^2\, \frac{d\Delta_{\rm disc}\hat\alpha}{d\mu^2} = \frac{\alpha}{\pi} \left( \sum_q Q_q \right)^2 \sigma,
\label{eq:relationozialpha}
\end{equation} 
so that,
\begin{equation}
\frac{d\tilde{\sigma}}{d\mu^2} = - \frac{\pi}{\alpha}\, \lambda_4 
\left( \sum_q Q_q \right)^{-2} \frac{d\Delta_{\rm disc}\hat\alpha}{d\mu^2} =
- \lambda_1\, \frac{\pi}{\alpha}\, \frac{d\Delta_{\rm disc}\hat\alpha}{d\mu^2}\ ,
\end{equation}
where the last step applies for $\mu < \bar m_c$ (we are assuming approximate isospin symmetry which 
eliminates the interval $\bar m_u < \mu < \bar m_d$).
Then,
\begin{equation}
\label{eq:relationozialpha2}
\tilde{\sigma}(\mu) - \tilde{\sigma}(\mu_{0}) = - \lambda_1\, \frac{\pi}{\alpha}\, 
[\Delta_{\rm disc}\hat{\alpha}(\mu) - \Delta_{\rm disc}\hat{\alpha}(\mu_{0})]. 
\end{equation}
These relations are general, but there is a subtle point. 
In general, the singlet pieces effectively decouple at renormalization scales $\bar m^{\rm disc}_q$
that may differ from the scales $\bar m_q$ at which the non-singlet pieces decouple.
This would generate various energy intervals with generally different values for $\lambda_1$. 
Implementing strong isospin symmetry in the form $\bar m_u = \bar m_d$ and 
$\bar m^{\rm disc}_u = \bar m^{\rm disc}_d$, as well as accepting the physical mass orderings 
$\bar m_s \geq \bar m_u$ and $\bar m^{\rm disc}_s \geq \bar m^{\rm disc}_u$, there remain a total of six 
different orderings. 

As an example, consider the case,
\begin{equation}
\bar m^{\rm disc}_s > \bar m_s > \bar m_u > \bar m^{\rm disc}_u.
\end{equation}
For scales $\mu > \bar m^{\rm disc}_s$ there are three active quarks with $Q_{u}+Q_{d}+Q_{s} = 0$ and
the singlet contributions vanish. 
For scales in the range $\bar m^{\rm disc}_s > \mu > \bar m_s$ we obtain the value $\lambda_{1}=1/2$.
Similarly, for $\bar m_s > \mu > \bar m_u$ and for $\bar m_u > \bar m^{\rm disc}_u$ we find 
$\lambda_1 = 9/20$ and 1/4, respectively. 
Below $\bar m^{\rm disc}_u$ all singlet contributions vanish by definition. 
Inserting these results into Eq.~(\ref{eq:relationozialpha2}) and summing the contributions from all 
intervals, we find the constraint, 
\begin{equation}
- \frac{\Delta_{\rm disc}\hat\alpha}{4} < \frac{\alpha}{\pi} \left[ \tilde\sigma(\bar m^{\rm disc}_s) 
- \tilde\sigma(\bar m^{\rm disc}_u) \right] < - \frac{\Delta_{\rm disc}\hat\alpha}{2}\ ,
\label{eq:oziinequality}
\end{equation}
where we have anticipated that $\Delta_{\rm disc}\hat\alpha < 0$ (see below). 

The other five cases are dealt with in the same way, 
and one can check that the inequality~(\ref{eq:oziinequality}) is never violated.
For the three mass orderings satisfying $\bar m^{\rm disc}_u \geq \bar m_u$, or generally if we can neglect 
the presumably small range $\bar m_u > \mu > \hat m^{\rm disc}_u$, we find the much stronger constraint,
\begin{equation}
- \frac{9 \Delta_{\rm disc}\hat\alpha}{20} < 
\frac{\alpha}{\pi} \left[ \tilde\sigma(\bar m^{\rm disc}_s) - \tilde\sigma(\bar m^{\rm disc}_u) \right] 
< - \frac{\Delta_{\rm disc}\hat\alpha}{2}\ .
\label{eq:oziinequality2}
\end{equation}

Since we do not expect the $\bar m^{\rm disc}_q$ to be numerically very different from the $\bar m_q$ we 
choose our central value to correspond to $\bar m^{\rm disc}_q = \bar m_q$, and we include twice the range in
Eq.~(\ref{eq:oziinequality2}) as the uncertainty due to possible $\bar m^{\rm disc}_q \neq \bar m_q$ effects.
Thus,
\begin{equation}
\label{eq:singlet}
\frac{\alpha}{\pi}
\left[ \tilde{\sigma}(\bar m^{\rm disc}_s) - \tilde{\sigma}(\bar m^{\rm disc}_u) \right] = 
- \left[ \frac{9}{20} \pm \frac{1}{20} \right] \, \Delta_{\rm disc}\hat\alpha,
\end{equation}
which can be inserted into Eq.~(\ref{eq:MASTEREQUATION}).
Notice, however, that Eq.~(\ref{eq:MASTEREQUATION}) also contains an implicit singlet contribution from each 
of the two terms in the first line.
Taken together, the $\lambda_1$ term cancels exactly the central value in Eq.~(\ref{eq:singlet}) and we finally arrive at
\begin{equation}
\Delta_{\rm disc}\hat s^2 = \left[ \hat s^2 \pm \frac{1}{20} \right] \Delta_{\rm disc}\hat\alpha.
\label{eq:relationoozi3}
\end{equation}

\begin{figure}[t]
\centering
\includegraphics[scale=0.40]{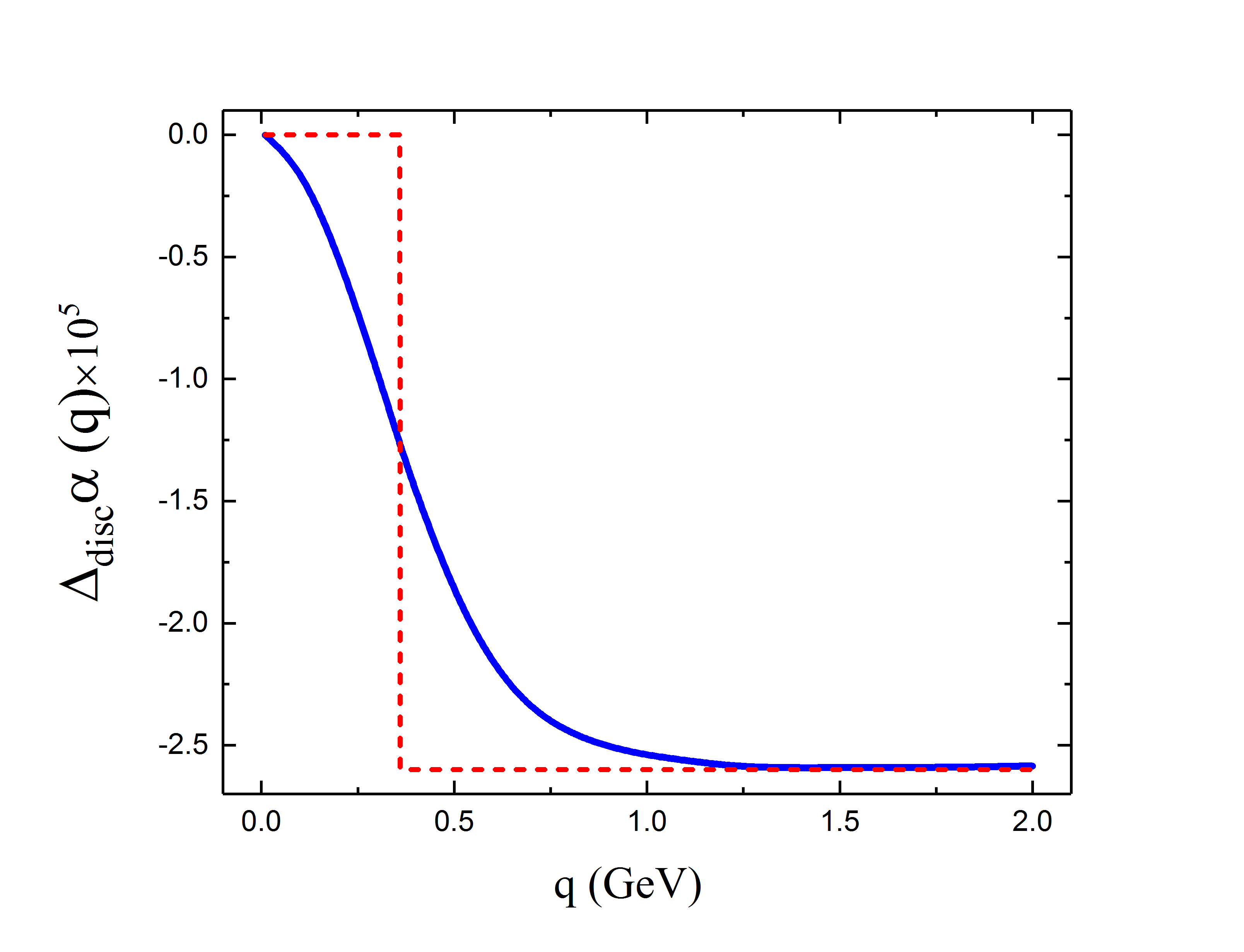}
\caption{Scale dependence of the singlet contribution to $\Delta\alpha$ (solid line) 
and its step function approximation (dashed line).}
\label{fig:OZI}
\end{figure} 

In Appendix~\ref{appendixA} we compute $\Delta_{\rm disc}\alpha$ in the on-shell scheme by exploiting the lattice gauge theory 
calculation~\cite{Blum:2015you} of the corresponding contribution to $a_\mu$ with the result,
\begin{equation}
\Delta_{\rm disc}\alpha(2.0\mbox{ GeV}) = -2.6 \times 10^{-5}.
\label{eq:OZIalpha}
\end{equation}
Note that because the sum of the charges of the three light quarks vanishes, and we enter the perturbative domain 
where the singlet piece is known to be tiny, we expect an asymptotically stable value at higher energies for 
$\Delta_{\rm disc}\alpha(q)$.  
This is supported by Figure~\ref{fig:OZI}, showing that $\Delta_{\rm disc}\alpha(q)$ is nearly $q$-independent 
for $q \gtrsim 1.2$~GeV.
We also remark that the dominance of low scales notwithstanding, the sign in Eq.~(\ref{eq:OZIalpha}) coincides 
with that of the singlet piece in the perturbative regime.
Also shown in Figure~\ref{fig:OZI} is the step function approximation of $\Delta_{\rm disc}\alpha(q)$, with the 
step defined as the value of $q$ where it reaches half of its asymptotic value in Eq.~(\ref{eq:OZIalpha}).
We interpret this as the value where the strange quark decouples from singlet diagrams, so that 
$\bar m^{\rm disc}_s \sim 350$~MeV.
Our central value of $\bar m_s$ to be derived in the next section, $\bar m_s = 342$~MeV, is numerically very 
close to this providing evidence for $\bar m^{\rm disc}_s \approx \bar m_s$.

Eq.~(\ref{eq:relationoozi3}) and Eq.~(\ref{eq:OZIalpha}) refer to quantities in the $\msbar$ and on-shell schemes,
respectively, and in general these may differ. 
However, since we are working here in the three quark theory and the sum of the charges of three light quarks vanishes, 
the change of schemes is trivial.  
We can therefore use Eq.~(\ref{eq:OZIalpha}) in Eq.~(\ref{eq:relationoozi3}) and obtain,
\begin{equation} 
\Delta_{\rm disc}\hat s^2 = (-0.6 \pm 0.3) \times 10^{-5},
\label{OZIfinalweak}
\end{equation}
where the uncertainty combines the errors from Eq.~(\ref{eq:relationoozi3}) and the one induced by the lattice 
calculation~\cite{Blum:2015you}.

\section{Flavor separation~\label{sec:flavor}}

\begin{table}[t]
\centering

\begin{tabular}{|lcrr|}
\hline 
\rule{-2pt}{3ex} channel & \phantom{OO} & $\ensuremath{a_{\mu}\times10^{10}}$ & 
\phantom{OOOOO}$\ensuremath{\Delta\alpha\times10^{4}}$\tabularnewline
\hline \hline
\rule{-2pt}{3ex} $\phi$ &  & 38.43 & 5.13\tabularnewline
$K\bar{K}\pi$ &  & 2.45 & 0.78\tabularnewline
$\eta\phi$ &  & 0.36 & 0.13\tabularnewline
PQCD~\cite{Erler:2000nx} ($>1.8\,\text{GeV}$) &  & 7.30 & --- \phantom{}\tabularnewline
\hline 
\rule{-2pt}{3ex} Total &  & 48.54 & 6.04\tabularnewline
\hline \hline
\rule{-2.8pt}{3ex} $K\bar{K}$ $(\text{non}-\phi)$ &  & 3.62 & 0.76\tabularnewline
$K\bar{K}2\pi$ &  & 0.85 & 0.30\tabularnewline
$K\bar{K}3\pi$ &  & -0.03 & -0.01\tabularnewline
$K\bar{K}\eta$ &  & 0.01 & 0.00\tabularnewline
$K\bar{K}\omega$ &  & 0.01 & 0.00\tabularnewline
\hline 
\rule{-2pt}{3ex} Total &  & 4.46 & 1.05\tabularnewline
\hline 
\end{tabular}
\caption{Channels associated with the strange quark external current (top) and possible further channels 
originating from it (bottom).
\label{tab:strangepossiblecontributions}}
\end{table}
In this section we perform a flavor separation of the contributions of up-type from down-type 
quarks, or --- given that up and down quarks are linked by the approximate strong isospin symmetry ---
a separation of $s$ from $u$ and $d$ quarks.
Our strategy consists of first using exclusively the experimental electro-production data as tabulated 
in Ref.~\cite{Davier:2017zfy} to constrain the contribution $\Delta_s\alpha$ of the strange quark to $\Delta\alpha$.
We then exploit the lattice gauge theory results in Refs.~\cite{Blum:2016xpd,Chakraborty:2014mwa} 
to confirm and refine the purely data driven analysis. 
Then we introduce the {\em threshold mass\/} $\bar{m}_q$ of a quark $q$ as the value of the 't Hooft 
scale where the QCD contribution to the corresponding decoupling relation becomes trivial. 
$\bar{m}_c$ and $\bar{m}_b$ are treated in perturbation theory, while for $u$, $d$, and $s$ quarks we 
derive bounds using phenomenological and theoretical constraints.  
\subsection{Experimental data}
To obtain $\Delta_s \alpha$ we use Ref.~\cite{Davier:2017zfy} where the contribution of each hadronic 
channel to $a_{\mu}$ and $\Delta\alpha$ for energies up to $1.8\,\text{GeV}$ is given. 
The main idea is to determine for each channel whether it was produced by an $\bar{s}s$ or 
a first generation quark current. 
For reasons that will become clear later, we consider both, $\Delta_s \alpha$, 
and the strange quark contribution to the anomalous magnetic moment, $a^s_{\mu}$.

We begin by listing in the upper part of Table~\ref{tab:strangepossiblecontributions} the experimental 
channels~\cite{Davier:2017zfy} which we associate with an $s\bar s$ current.  
Up to OZI-rule violating $\phi$-$\omega$ and $\phi$-$\rho$ mixing effects, 
the $\phi$~meson can be identified with strange quarks. 
We calculate its contribution using a Breit-Wigner shape with $s$-dependent total and partial widths, adopting 
the PDG values~\cite{Patrignani:2016xqp} for the $\phi$ meson branching ratios and applying a small correction 
for $\phi$-$\omega$ mixing.
As for the $\phi(1680)$, the main decay channel is $K\bar{K^{*}}$ with $K^{*}$ mesons decaying almost 
entirely into $K\pi$.
As can be seen from data~\cite{Davier:2017zfy}, the $K\bar{K}\pi$ channel is indeed virtually saturated by 
$\phi(1680)$ decays.
The $\eta$-$\phi$ channel also arises dominantly from the strange quark current since the contribution to this 
channel from light quarks is Zweig rule suppressed. 
Conversely, we expect channels involving an $\eta$ meson accompanied by non-strange states to be mainly due to 
light quark currents. For $a^s_\mu$ we need to add the contribution from energies above 1.8~GeV. 
It can be computed within PQCD and taken as one sixth 
of the corresponding light quark contribution~\cite{Erler:2000nx} of $43.8\times 10^{-10}$. 
The lower part of Table~\ref{tab:strangepossiblecontributions} shows further channels involving strange quarks 
to which first generation quark currents could conceivably contribute, and we conservatively assign 
$(50 \pm 50)$\% of these to the $s\bar s$ current. 
The table also shows the corresponding contributions to $a_{\mu}$.
Adding the totals in this way we find,
\begin{equation}
a^s_\mu = (50.77 \pm 0.60 \pm 0.83 \pm 2.23) \times 10^{-10} = (50.77 \pm 2.45) \times 10^{-10},
\label{eq:afirst}
\end{equation}
and,
\begin{equation}
\Delta_s\alpha (1.8\mbox{ GeV}) = (6.56 \pm 0.11 \pm 0.19 \pm 0.53)\times 10^{-4} = (6.56 \pm 0.57)\times 10^{-4}.
\label{eq:alphafirst}
\end{equation}
The first errors are experimental~\cite{Davier:2017zfy} where we accounted for correlations.
The second errors allow for differences in parametrizations when decay parameters are extracted from 
experimental data by different groups.
The last errors are half of the totals in Table~\ref{tab:strangepossiblecontributions}, but we expect the 
$s\bar s$ current to virtually saturate the kaon channels in Table~\ref{tab:strangepossiblecontributions}
because the larger strange quark mass should suppress the probability amplitude to produce an $s\bar s$ sea 
quark pair relative to first generation quark pairs. 

The uncertainty in Eq.~(\ref{eq:alphafirst}) is already about three times smaller than in the past~\cite{Erler:2004in}.
We can reduce it further by quantifying our expectation that the strange quark current actually saturates the 
kaon channels listed in the bottom part of Table~\ref{tab:strangepossiblecontributions}. 
For this, we re-write Eqs.~(\ref{eq:afirst}) and (\ref{eq:alphafirst}) in the form,
\begin{eqnarray}
a^s_\mu &=& (53.00 - 4.46\, \kappa \pm 0.60 \pm 0.83) \times 10^{-10},
\label{eq:dataa} \\[6pt]
\Delta_{s}\alpha (1.8\mbox{ GeV}) &=& (7.09 - 1.05\, \kappa \pm 0.11 \pm 0.19) \times 10^{-4},
\label{eq:datadelta}
\end{eqnarray} 
with a parameter $0 \leq \kappa \leq 1$, where $\kappa = 0$ ($\kappa = 1$) corresponds to the case where all kaon 
contributions in Table~\ref{tab:strangepossiblecontributions} arise from the strange (first generation) quark current. 
In order to confirm that indeed $\kappa \approx 0$ and to compute an uncertainty for possible $\kappa \neq 0$ 
effects, we can use results on $a^s_\mu$ from lattice gauge theory, as we show next.

\subsection{Lattice data}
Two groups~\cite{Blum:2016xpd,Chakraborty:2014mwa} calculated the contribution of the strange quark to the 
vacuum polarization function within lattice gauge theory with a focus on $a^s_\mu$. 
The two results agree and average to
\begin{equation}
a^s_\mu = (53.32 \pm 0.49) \times10^{-10} \quad [{\rm lattice}],
\label{latticeamu}
\end{equation}
which is in perfect agreement with Eq.~(\ref{eq:dataa}) and our expectation $\kappa \approx 0$.
Since the analogous result for $\Delta_{s}\alpha (1.8\mbox{ GeV})$ has not been provided by either of the 
groups, we follow a Bayesian procedure to quantify the parameter $\kappa$ in Eq.~(\ref{eq:datadelta}), using 
as prior information the comparison of Eq.~(\ref{eq:dataa}) with Eq.~(\ref{latticeamu}).
The 68.3\% highest probability interval of $\kappa$, namely $0 \leq \kappa \leq \kappa_{1\sigma}$, 
can be obtained from
\begin{equation}
N\int\limits^{\kappa_{1\sigma}}_{\phantom{O}0} \exp\left[ 
-\frac{(53.00 - 4.46\, \kappa - 53.32)^2}{2(0.60^2 + 0.83^2 + 0.49^2)} \right] d\kappa = 0.683,
\end{equation}
where $N$ is the normalization of the distribution. 
This yields $\kappa_{1\sigma} = 0.22$, and Eq.~(\ref{eq:datadelta}) now provides us with the desired result,
\begin{equation}
\Delta_s\alpha (1.8\mbox{ GeV}) = (7.09 \pm 0.11 \pm 0.19 \pm 0.23)\times10^{-4} = (7.09 \pm 0.32) \times 10^{-4},
\label{eq:datadelta1}
\end{equation}
which is consistent with, but more precise than Eq.~(\ref{eq:alphafirst}).
We assigned the uncertainty from $\kappa_{1\sigma}$ symmetrically around $\kappa = 0$, which is both the 
physically favored and most probable value (the peak of the distribution). 
This rather conservative treatment effectively doubles the error from $\kappa_{1\sigma}$,
and is meant to account for the fact that the kernels of $\Delta\alpha$ and $a_\mu$ differ.

The experimental values of $a_{\mu}$ and $\Delta\alpha$ are correlated, 
possibly impacting Eq.~(\ref{eq:datadelta1}).
However, we found that even assuming them to be fully correlated changes the central value only very slightly 
and reduces the uncertainty modestly.
Thus, we keep Eq.~(\ref{eq:datadelta1}) as our final result on $\Delta_s\alpha (1.8\mbox{ GeV})$.
As an additional cross-check we used the vacuum polarization function of another lattice 
calculation~\cite{Chakraborty:2014mwa} of $a_\mu^s$ (expressed as a Pad\'e approximant which is the source of 
the largest uncertainty~\cite{Chakraborty:2014mwa}) to first reproduce their results, and then we computed 
$\Delta_{s}\alpha$ which yields,
\begin{equation}
\Delta_s\alpha  (1.8\,\text{GeV}) \approx(6.9\pm0.5)\times10^{-4} \quad [{\rm lattice}],
\label{eq:datadelta1latticce}
\end{equation} 
in excellent agreement with Eq.~(\ref{eq:datadelta1}).

\subsection{Threshold masses}
\subsubsection{Heavy quarks}
We can compute $\bar{m}_c$ and $\bar{m}_b$ in perturbation theory by reincorporating the RGE summable 
logarithms of the form $\ln \hat m_q/\bar m_q$ into Eq.~(\ref{eq:PIo}), and then solving for $\bar m_q$ by 
setting the contribution from quark $q$ equal to zero.  
Since $\bar m_q \to \hat m_q$ for $\hat\alpha_s \to 0$, these logarithms are at most of order $\hat\alpha_s$ 
and can be ignored in the $\hat\alpha_s^3$ coefficient. 
Thus, we can use a previous analysis~\cite{Erler:1998sy} where the logarithms up to order $\hat\alpha_s^2$ are given.
We find,
\begin{eqnarray} 
\nonumber
\bar m &=& \hat m  \left\{1- \frac{13}{24} \frac{\hat\alpha_s}{\pi} + \left( \frac{10073}{3456} - 
\frac{655}{288} \zeta_3 - \frac{361}{2592} n_q \right) \frac{\hat\alpha_s^2}{\pi^2} \right. \\[12pt]
\nonumber &+& \left( 1.61024 + 0.59599\, n_q - 0.00964\, n_q^2 \right) \frac{\hat\alpha_s^3}{\pi^3} \\[12pt]
&+& \left.\left[ - \frac{295}{2592} \frac{\hat\alpha_s^2}{\pi^2} + \left( \frac{5767}{62208} - 
\frac{\mathcal K_1 + \mathcal K_2 n_q}{2} \right) \frac{\hat\alpha_s^3}{\pi^3} \right] 
\frac{\sum Q_\ell^2}{Q_h^2} \right\}. 
\end{eqnarray}
Using the input values for the $Z$ boson mass~\cite{Patrignani:2016xqp}, $M_Z = 91.1876$~GeV, the charm quark 
mass~\cite{Erler:2016atg}$, \hat m_c(\hat m_c) = 1.272$~GeV, and the bottom quark 
mass~\cite{Patrignani:2016xqp}, $\hat m_b(\hat m_b) = 4.18$~GeV, together with the 4-loop 
RGE~\cite{vanRitbergen:1997va} for $\hat\alpha_s$ with $n_q = 4$ and $n_q=5$, respectively, we find 
\begin{eqnarray}
\bar{m}_c &=& 1.185 \mbox{ GeV}, 
\label{eq:mcbar} \\[6pt]
\bar m_b &=& 3.990 \mbox{ GeV}. 
\label{eq:mbbar}
\end{eqnarray}

It will be useful for later to define quantities $\xi_{q}$~\cite{Erler:2004in} as ratios between the 
threshold mass of quark $q$ and the $1S$ $\bar{q}q$ bound state mass, 
\begin{equation}
\xi_{q}\equiv\frac{2\bar{m}_{q}}{M_{1S}}\ .
\end{equation}
This definition implies that $\xi_{q}\rightarrow 1$ for $\bar{m}_{q}\rightarrow\infty$ and
$\xi_{q}\rightarrow 0$ for $\bar{m}_{q}\rightarrow0$. 
We expect $\xi_q$ to be a monotonically increasing in the sense that $\xi_1 > \xi_2$ if 
$\bar m_1 > \bar m _2$. 
Using the PDG values for the bound state masses~\cite{Patrignani:2016xqp} we find $\xi_c = 0.766$ and 
$\xi_b = 0.844$, and thus $\xi_{b}>\xi_{c}$ as expected.

\subsubsection{Light quarks}
Next we constrain the individual contributions of the light quarks to $\Delta\hat\alpha$, evaluated at $\bar m_c$. 
Using the RGE and the starting value given in Eq.~(\ref{eq:deltatwogevms}) we obtain,
\begin{equation}
\Delta\hat\alpha^{(3)}(\bar{m}_c) = (65.10 \pm 0.45 \pm 0.18) \times 10^{-4}.
\label{eq:deltaalgamcbar}
\end{equation}
From Eq.~(\ref{eq:datadelta1}) we can also calculate $\Delta_{s}\hat{\alpha}$ at $\bar{m}_{c}$. 
To do so, we first invoke experimental data to obtain the shift,
\begin{equation}
\Delta_s\alpha (2\mbox{ GeV}) = \Delta_s\alpha (1.8\mbox{ GeV}) + (0.55 \pm 0.04) \times10^{-4},
\end{equation}
given by one sixth of the continuum contribution~\cite{Davier:2017zfy} of $(3.31 \pm 0.26) \times10^{-4}$ 
between the two scales. 
The uncertainty is the difference to using PQCD instead of data and accounts for quark-hadron duality violations.
Changing to the $\msbar$ scheme and employing again the RGE gives,
\begin{equation}
\Delta_s \hat\alpha(\bar m_c) = (8.71 \pm 0.32) \times 10^{-4}.
\label{eq:datadelta1mc}
\end{equation} 
Since the threshold mass is the value of the 't~Hooft scale corresponding to trivial matching conditions
regarding the QCD contribution, we can write,
\begin{equation}
\Delta_s\hat\alpha(\bar m_c) = Q_s^2\frac{\alpha}{\pi} K_{\rm QCD}^s (\bar m_c) \ln\frac{\bar m_c^2}{\bar m_s^2}\ ,
\label{eq:alphas}
\end{equation}
where we defined a scale dependent factor $K_{\rm QCD}^q(\mu)$ as the average QCD correction to the $\beta$ 
function between $\bar m_q$ and the scale $\mu$. 
Eq.~(\ref{eq:alphas}) has two unknowns, $K_{\rm QCD}^s (\bar m_c)$ and $\bar m_s$, and it shows that 
increasing $K_{\rm QCD}^s (\bar m_c)$ forces the logarithm to decrease and in turn $\bar m_s$ to increase. 
Thus, smaller (larger) values of $K_{\rm QCD}^s(\bar m_c)$ correspond to a smaller (larger) values of $\bar m_s$. 
On the other hand, if we have two quarks with masses $\bar m_1 > \bar m_2$, we expect the average QCD contribution between 
$\bar m_2$ and $\mu$ to be larger than that between $\bar m_1$ and $\mu$, since $\alpha_s$ is larger at lower scales. 
Thus,
\begin{equation}
\bar m_1 > \bar m_2 \qquad \implies \qquad K_{\rm QCD}^1(\mu) < K_{\rm QCD}^2(\mu),
\end{equation}
and we must have, 
\begin{equation}
K^c_{\rm QCD}(\bar m_c) < K^s_{\rm QCD}(\bar m_c).
\end{equation}
$K_{\rm QCD}^c(\bar m_c)$ can be computed from Eq.~(\ref{eq:ki}). 
Using $n_q = 3$ and $\alpha_s(\bar m_c) = 0.413$ yields $K_{\rm QCD}^c(\bar m_c)=1.178$, and implies the lower bound,
\begin{equation}
\bar m_s > 
\bar m_c \exp \left[ - \frac{\pi\Delta_s\hat\alpha(\bar m_c)}{2 Q_s^2 \alpha K_{\rm QCD}^c(\bar m_c)} \right] 
= 289\mbox{ MeV},
\end{equation}
where we used $\alpha = \alpha(\bar m_s) \approx 1/135$.
We can also obtain an upper bound on $\bar m_s$,
\begin{equation}
\frac{2\bar m_s}{M_\phi} = \xi_s < \xi_c = 0.766 \qquad \implies \qquad \bar m_s < 390\mbox{ MeV},
\end{equation}
implying $K^s_{\rm QCD}(\bar m_c) < 1.50$.
We can summarize these results by writing,
\begin{equation} 
K^s_{\rm QCD}(\bar m_c) = 1.34 \pm 0.16, \qquad\qquad \bar m_s = 342_{-53}^{+48}\mbox{ MeV}.
\label{eq:kqcd1}
\end{equation}
$\bar m_u$ and $\bar m_d$ can be obtained in a similar way.
We have,
\begin{equation}
\Delta_{\rm conn}\hat\alpha^{(3)}(\bar m_c) = \Delta_s\hat\alpha(\bar m_c) + \frac{2\alpha}{\pi} 
\left[ (Q_u^2 + Q_d^2)K_{\rm QCD}^{u,d} \ln\frac{\bar m_c}{\bar m_{u,d}}\right],
\end{equation}
where the quark connected contribution to $\Delta\hat\alpha^{(3)}(\bar m_c)$ is given by,
\begin{equation}
\Delta_{\rm conn}\hat\alpha^{(3)}(\bar m_c) \equiv
\Delta\hat\alpha^{(3)}(\bar m_c) - \Delta_{\rm disc}\hat\alpha^{(3)}(\bar m_c) = 65.36\times 10^{-4}.
\end{equation}
Following the same steps as for $\bar m_s$ we find,
\begin{equation}
K^{u,d}_{\rm QCD}(\bar m_c) = 1.38 \pm 0.20, \qquad\qquad \bar m_{u,d} = 246_{-57}^{+54}\mbox{ MeV},
\label{eq:kqcd2}
\end{equation}
where the errors in Eqs.~(\ref{eq:kqcd1}) and (\ref{eq:kqcd2}) are strongly correlated.
The light quark threshold masses are convenient for implementing the RGE and serve an illustrative purpose, 
but their precise values affect $\hat s(0)$ only at order ${\cal O}(\alpha^2)$ and beyond, as long as the 
central value in Eq.~(\ref{eq:datadelta1mc}) remains fixed (the uncertainty there will give rise to the flavor 
separation error).
Notice, that for the central values we have $\bar m_s - \bar m_u \approx 96$~MeV, which is of typical size for 
hadronic mass splittings within $SU(3)$ flavor multiplets.

Finally, accounting for the squares of the electric charges we obtain the contributions from the first 
generation quarks at the scale $\bar{m}_{s}$,
\begin{equation}
\Delta\hat\alpha^{(2)}(\bar m_s) = \Delta\hat\alpha^{(3)}(\bar m_c) - 6 \Delta_s\hat\alpha(\bar m_c)
= (12.9 \mp 1.9) \times10^{-4},
\label{eq:relation2andthree}
\end{equation}
where we only quote the uncertainty from the flavor separation in Eq.~(\ref{eq:datadelta1mc}).

\section{Theoretical uncertainties~\label{sec:uncertainties}} 
In addition to parametric uncertainties, there are five sources of theoretical uncertainties for the weak 
mixing angle at low energies affecting our calculation. 
\begin{table}[t]
\centering
\begin{tabular}{|lcr|}
\hline 
\rule{0pt}{3ex} source & \phantom{OO} & $\ensuremath{\delta\sin^{2}\hat{\theta}_{W}(0)\times10^{5}}$\tabularnewline
\hline 
\rule{-3.5pt}{3ex} $\ensuremath{\Delta\hat{\alpha}^{(3)}(2\mbox{ GeV})}$ &  & 1.2\phantom{O}\tabularnewline
flavor separation &  & 1.0\phantom{O}\tabularnewline
isospin breaking &  & 0.7\phantom{O}\tabularnewline
singlet contribution &  & 0.3\phantom{O}\tabularnewline
PQCD &  & 0.6\phantom{O}\tabularnewline
\hline 
\rule{0pt}{2.5ex}Total &  & 1.8\phantom{O}\tabularnewline
\hline 
\end{tabular}
\caption{Theoretical uncertainties in the low energy mixing angle.}
\label{tab:errors}
\end{table}
They are summarized in Table~\ref{tab:errors} and discussed in the following.

The first uncertainty is induced by the experimental error in the determination of $\Delta\hat\alpha^{(3)}(2.0\mbox{ GeV})$.
Eq.~(\ref{eq:MASTEREQUATION}) propagates this uncertainty to the weak mixing angle~\cite{Erler:2004in}, 
\begin{equation}
\delta\hat s^2(0) = \left[ \frac{1}{2} - \hat s^2 \right] 
\delta\Delta\hat\alpha^{(3)}(2\mbox{ GeV}) = \mp 1.2 \times10^{-5},
\label{eq:uncertaintyalpha}
\end{equation}
where we have used $\delta\Delta\hat\alpha^{(3)}(2\mbox{ GeV}) = \pm 0.45 \times10^{-4}$ from Eq.~(\ref{eq:noglobalfit}).

The three light quarks enter with different electroweak weights into $\hat s^2(0)$ and $\Delta\alpha^{(3)}(\bar m_c)$.
The flavor separation uncertainty is due to the imperfect knowledge of how much $s$ quarks relative to $u$ and 
$d$ quarks contribute to $\Delta\alpha^{(3)}(\bar{m}_{c})$.
It is given by~\cite{Erler:2004in},
\begin{equation}
\delta\hat s^2(0) \simeq \frac{1}{20}\, \delta\Delta\hat\alpha^{(2)}(\bar m_c) = \pm 1.0 \times10^{-5},
\label{eq:uncertantystrange}
\end{equation}
where we used $\delta\Delta\hat\alpha^{(2)}(\bar m_s) = \pm 1.9 \times10^{-4}$ from Eq.~(\ref{eq:relation2andthree}).

The flavor separation assumed isospin symmetry in the form $\bar m_u = \bar m_d$. 
To estimate the uncertainty associated with isospin breaking, we first consider the idealized case in which 
$SU(2)$ isospin violation was as large as $SU(3)$ breaking.
This would occur for $\bar m_d = \bar m_s$, so that from Eq.~(\ref{eq:relation2andthree}) the $u$ quark 
current could at most contribute
\begin{equation}
\Delta\alpha^{(1)}(\bar m_d) < 14.8 \times10^{-4}.
\end{equation}
To propagate this uncertainty to $\hat s^2(0)$ we can use~\cite{Erler:2004in},
\begin{equation}
\delta \hat s^2(0) = - \frac{3}{40} \Delta\alpha^{(1)}(\bar m_d) > -1.1 \times10^{-4}.
\end{equation}
A measure of the breaking of $SU(2)$ relative to $SU(3)$ is given by the ratio,
\begin{equation}
\left| \frac{M_{K^{*\pm}}^2 - M_{K^{*0}}^2}{M_{K^{*\pm}}^2 - M_{\rho^0}^2} \right| \approx 0.06,
\end{equation}
so that,
\begin{equation}
\delta\hat s^2(0) = ^{+0}_{-7} \times 10^{-6}.
\end{equation}
This error is asymmetric because we assume $\bar m_d \geq \bar m_u$, but it is convenient and conservative to 
treat it symmetrically in Table~\ref{tab:errors}.

The uncertainty arising from the singlet contribution is given in Eq.~(\ref{OZIfinalweak}).
The last entry in Table~\ref{tab:errors} combines the truncation error from the perturbative matching 
conditions with the scheme conversion error shown as the second uncertainty in Eq.~(\ref{eq:deltatwogevms}).

\section{Results and conclusions~\label{sec:conclusions}} 
Eq.~(\ref{eq:MASTEREQUATION}) together with the $Z$ pole value of the weak mixing angle from a global fit 
to the SM~\cite{Patrignani:2016xqp}, $\sin^2\hat\theta_W (M_{Z}) = 0.23129(5)$, can now be used to compute the weak 
mixing angle at zero momentum transfer,
\begin{equation}
\sin^{2}\hat{\theta}_{W}\left(0\right)= 0.23868 \pm 0.00005 \pm 0.00002,
\label{eq:weakanglenoglobal}
\end{equation}
where the second error is the total theoretical uncertainty from Table~\ref{tab:errors}.

\begin{figure}[t]
\centering
\includegraphics[scale=0.45]{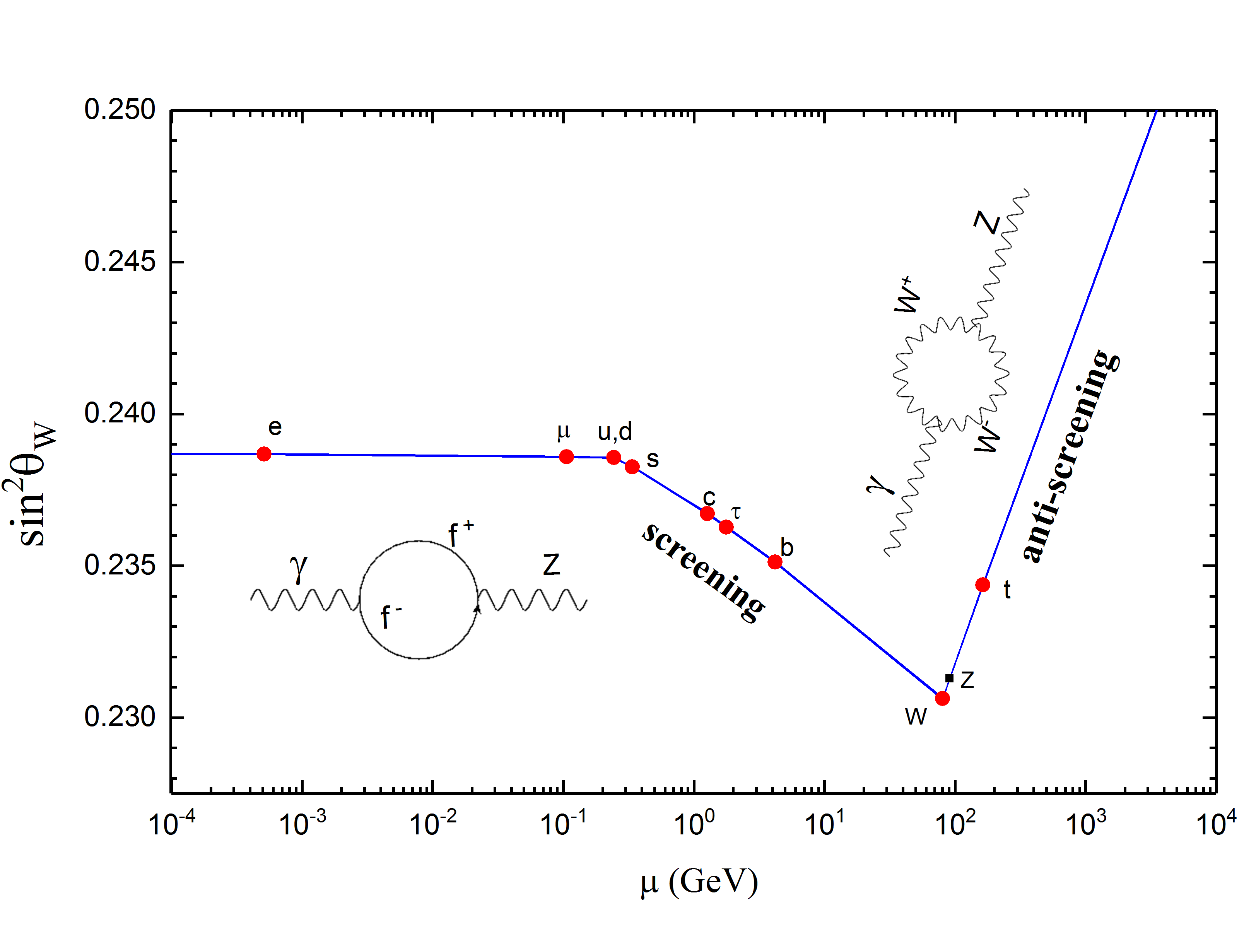}
\caption{Scale dependence of the weak mixing angle in the $\msbar$ renormalization scheme.
The dots indicate the scales where a particle is integrated out.  
The total uncertainty corresponds to the thickness of the line.
The $\beta$-function of $SU(2)_L$ changes sign at $\mu = M_W$, where the fermionic screening effects of the 
effectively Abelian gauge theory are being overcompensated by the anti-screening effects of the full 
non-Abelian electroweak theory.}
\label{fig:WeakAngle}
\end{figure}

To facilitate the update of our results in the future, we also present a linearized formula of the form factor $\kappa(0)$,
\begin{equation}
\sin^2\hat\theta_W(0) \equiv \hat\kappa(0)\sin^2\hat\theta_W(M_{Z}),
\label{eq:kappadefinition}
\end{equation}
in terms of 
variations of the input parameters, using $\Delta\hat\alpha_s(M_Z)$ in Eq.~(\ref{eq:alphasvar}), as well as,
\begin{equation}
\tilde{\Delta}\alpha\equiv \Delta\alpha(2.0\mbox{ GeV}) - 0.005871,
\end{equation}
and,
\begin{equation}
\Delta\hat m_c \equiv \frac{\hat m_c(\hat m_c)}{1.272\mbox{ GeV}} - 1, \qquad\qquad
\Delta\hat m_b \equiv \frac{\hat m_b(\hat m_b)}{4.180\mbox{ GeV}} - 1.
\end{equation}
We obtain,
\begin{equation}
\hat{\kappa}(0) = 1.03196 \pm 0.00006 + 1.14\, \tilde\Delta\alpha + 0.025\, \Delta\hat\alpha_s
- 0.0016\, \Delta\hat m_c - 0.0012\, \Delta\hat m_b\ ,
\end{equation}
which shows that the current experimental uncertainties of $\pm 0.45 \times 10^{-4}$ 
in $\Delta\alpha(2\mbox{ GeV})$ from Eq.~(\ref{eq:relationoozi3}) and of $\pm 0.0016$ in $\hat\alpha_s(M_Z)$ 
induce errors of $\pm 5 \times 10^{-5}$ and $\pm 4 \times 10^{-5}$ in $\hat{\kappa}(0)$, respectively. 
Variations of $\pm 8$~MeV~\cite{Erler:2016atg} in $\hat m_c (\hat m_c)$ and $\pm 30$~MeV in 
$\hat m_b (\hat m_b)$ both imply $\mp 2 \times 10^{-6}$ in $\hat s(0)$ which is negligible. 
The resulting scale evolution of the weak mixing angle is illustrated in Figure~\ref{fig:WeakAngle}. 

When our result for the weak mixing angle in the Thomson limit or some other low momentum scale is used for 
the calculation of physical observables, there will generally be further process-dependent radiative 
corrections which need to be addressed. 
We expect this to be possible with theoretical uncertainties well below those in $\sin^2\hat\theta_W(0)$
summarized in Table~\ref{tab:errors}.
Thus, we reduced the total theoretical uncertainty in the weak mixing angle at low energies
from $7 \times 10^{-5}$~\cite{Erler:2004in} to less than $2 \times 10^{-5}$ which can safely be neglected for 
any current or planned experiment.

In summary, we developed a new way of calculating the flavor separation which involved 
both $e^+ e^- \rightarrow$ hadrons data and results from lattice gauge theory. 
We also better control now the uncertainty in the contribution of disconnected diagrams where 
we exploited results of Ref.~\cite{Blum:2015you} on the anomalous magnetic moment. 
Furthermore, we extended various formulas to the next order in perturbation theory,
reducing the perturbative uncertainty. 
There has also been significant progress in the evaluations of 
$\Delta\alpha$~\cite{Davier:2017zfy,Jegerlehner:2017zsb} and $\Delta\hat m_c(\hat m_c)$~\cite{Erler:2016atg}.
The theoretical uncertainty in $\sin^2\hat\theta_W(0)$ is now at a negligible level.

\appendix
\section{Calculations of \texorpdfstring{$\alpha(M_Z^2)$}{Lg}}
\label{appendixB}
Three independent groups presented recent evaluations of the hadronic contribution to the scale dependence of $\alpha$. 
In this appendix we briefly compare their approaches and results. 

In the Adler function approach~\cite{Eidelman:1998vc,Jegerlehner:2017zsb}, one uses the relations,
\begin{equation}
\frac{D(Q^2)}{Q^2} \equiv 12\pi^2\frac{d\Pi(q^2)}{dq^2} = 
-\frac{3\pi}{\alpha}\frac{d}{dq^2}\Delta_{\rm had}(q^2) = 
\int^{\infty}_{4m^2}\frac{R(s)}{(s+Q^2)^2}ds\ ,
\end{equation}
where $Q^2=-q^2$, and where the dispersion integral in the latter expression can be used to implement experimental data
up to some cut-off $M_0$. 
One can then write, 
\begin{eqnarray} 
\nonumber
\Delta_{\rm had}\alpha^{(5)}(M^2_Z) &=& \Delta_{\rm had}\alpha^{(5)}(-M^2_0)_{\rm data} + 
[\Delta_{\rm had}\alpha^{(5)}(-M^2_Z)-\Delta_{\rm had}\alpha^{(5)}(-M^2_0)]_{\rm PQCD} \\[12pt]
&+& [\Delta_{\rm had}\alpha^{(5)}(M^2_Z)-\Delta_{\rm had}\alpha^{(5)}(-M^2_Z)]_{\rm PQCD}\ , 
\end{eqnarray}
where the last two terms are computed using the operator product expansion (OPE) of $R(s)$, i.e., 
including the leading non-perturbative condensate corrections. 
Demanding consistency with the OPE of the Adler function itself suggests that a value of $M_0$ as low as 2~GeV appears to 
be a safe choice. 
Using this approach implies~\cite{Jegerlehner:2017zsb} for the on-shell definition,
\begin{equation}
\alpha(M^2_Z)^{-1} = 128.958 \pm 0.016\ .
\label{alphaFJ}
\end{equation}

The approach of Ref.~\cite{Hagiwara:2011af} is mostly data driven. 
Experimental data were used up to 11.09~GeV (except for the interval between 2.6~GeV and 3.73~GeV) and PQCD beyond that. 
The dispersion relation~(\ref{SDR}) then implied,
\begin{equation}
\label{alphaTeubner}
\alpha(M^2_Z)^{-1} = 128.944 \pm 0.019\ .
\end{equation}

Similarly, Ref.~\cite{Davier:2017zfy} uses data up to only 5~GeV (except for the interval between 1.8~GeV and 3.7~GeV), 
with the result,
\begin{equation}
\alpha(M^2_Z)^{-1} = 128.947 \pm 0.012\ .
\end{equation}
Here, we rely on the data handling of this work as it includes much more recent data than Ref.~\cite{Hagiwara:2011af}.
Moreover, the breakdown of individual channels and energy ranges is more explicit compared to
Ref.~\cite{Jegerlehner:2017zsb}.

Finally, changing our own result, with $\hat{\alpha}(M^2_{Z})^{-1} = 127.959 \pm 0.010$,  
based on the direct application of the renormalization group and matching equations and including $\tau$ 
decay data, from the $\msbar$ scheme to the on-shell scheme including the top quark contribution, 
we find,
\begin{equation}
\alpha(M^2_Z)^{-1} = 128.949 \pm 0.010\ .
\label{ouralpha}
\end{equation}
The numerical difference of our result to Ref.~\cite{Davier:2017zfy} arises mostly from the 
different\footnote{The three groups use slightly different values for $\alpha_{s}$, 
but this amounts to difference below the level of $0.004$ in $\alpha(M_Z)^{-1}$.} value of $\alpha_{s}$ 
and our treatment of the charm quark contribution~\cite{Erler:2016atg}. 
Thus, in view of the rather different approaches and differences in data sets, 
all numerical results are in good agreement with each other. 

\section{Calculation of \texorpdfstring{$\Delta_{\rm disc}\hat\alpha$}{Lg} }
\label{appendixA}
In the on-shell scheme one has~\cite{Bernecker:2011gh},
\begin{equation}
\Delta_{\rm disc}\alpha(q) = 4\pi\alpha\, {\rm Re} \left[ \Pi(q^2)-\Pi(0)\right]_{\rm disc}\ ,
\end{equation}
where,
\begin{equation}
\left[ \Pi(q^2)-\Pi(0)\right]_{\rm disc} = 
\sum_{t=0}^{T} \left[ \frac{\cos(qt)-1}{q^ 2} + \frac{t^2}{2} \right] C(t).
\end{equation}
$C(t)$ has been computed~\cite{Blum:2015you} in units set by the lattice cut-off scale $a^{-1} = 1.73$~GeV.
To obtain Eq.~(\ref{eq:OZIalpha}), we plotted $\Delta_{\rm disc}\alpha(q)$ as a function of $T$ and observe 
a plateau near $T = 20$, which closely mirrors the result for the case of $a_\mu$.
The value of the plateau is interpreted as the physical value~\cite{Blum:2015you}.
As an independent check we compute the ratio $\rho(a_\mu)$ of the disconnected contribution to the anomalous 
magnetic moment~\cite{Blum:2015you}, $a^{\rm disc}_\mu = -9.6\times 10^{-10}$, to the total hadronic 
contribution~\cite{Davier:2017zfy} for energies up to 1.8~GeV, obtaining $\rho(a_\mu) = -0.015$.  
The integration kernel of $a_\mu$ enhances contributions from low $q^2$ momenta, and recalling that 
$Q_u + Q_d + Q_s = 0$, the disconnected piece also predominantly arises from such momenta.
On the other hand, the integration kernel for $\Delta\alpha$ has greater support at higher scales compared 
to $a_\mu$, so that $\rho(a_\mu)$ should imply an upper bound on the disconnected contribution to 
$\Delta\alpha$. 
Numerically,
\begin{equation}
\left|\Delta_{{\rm disc}}\alpha\left(1.8\,\text{GeV}\right)\right| < 
\left|\rho\left(a_{\mu}\right)\times\Delta_{{\rm had}}\alpha\left(1.8\,\text{GeV}\right)\right| = 
8.3\times10^{-5},
\end{equation}
where $\Delta_{\rm had}\alpha(1.8\mbox{ GeV}) = 55.26 \times 10^{-4}$. 
This confirms the finding in Eq.~(\ref{eq:OZIalpha}) that $\Delta_{\rm disc}\alpha$ is very small.

\acknowledgments

This work is supported by CONACyT (Mexico) project 252167--F and the German--Mexican research collaboration 
grant SP~778/4--1 (DFG) and 278017 (CONACyT). 

\section*{Note added}
When this paper was under revision, a new analysis of $\alpha(M_Z^2)$ appeared~\cite{Keshavarzi:2018mgv}. 
Their value,
\begin{equation}
\alpha(M^2_Z)^{-1} = 128.946 \pm 0.015\ ,
\end{equation}
is even closer to our result~(\ref{ouralpha}) than the previous analysis in Ref.~\cite{Hagiwara:2011af}. 
However, there are some non-negligible differences in specific channels between Refs.~\cite{Davier:2017zfy}
and~\cite{Keshavarzi:2018mgv}.


\begin{thebibliography}{99}

\bibitem{Czarnecki:2000ic} 
A.~Czarnecki and W.~J.~Marciano, 
\emph{Polarized M\o ller scattering asymmetries}, \\
\emph{Int.\ J.\ Mod.\ Phys.}\ {\bf A15} (2000) 2365
[hep-ph/0003049].

\bibitem{Erler:2004in} 
J.~Erler and M.~J.~Ramsey-Musolf, 
\emph{The weak mixing angle at low energies}, \\
\emph{Phys.\ Rev.}\ {\bf D72} (2005) 073003
[hep-ph/0409169].

\bibitem{Androic:2013rhu} 
Qweak Collaboration: D.~Androic \etal,
\emph{First determination of the weak charge of the proton},
\emph{Phys.\ Rev.\ Lett.}\ {\bf 111} (2013) 141803
[arXiv:1307.5275].


\bibitem{Becker:2018ggl} 
D.~Becker \etal,
\emph{The P2 Experiment --- A future high-precision measurement of the electroweak mixing angle at low 
momentum transfer}, [arXiv:1802.04759].

\bibitem{Benesch:2014bas} 
MOLLER Collaboration: J.~Benesch \etal,
\emph{The MOLLER experiment: an ultra-precise measurement of the weak mixing angle using M\o ller 
scattering}, \\
\emph{Jefferson Lab document}, JLAB--PHY--14--1986
[arXiv:1411.4088].

\bibitem{Anthony:2005pm} 
SLAC--E158 Collaboration: P.~L.~Anthony \etal,
\emph{Precision measurement of the weak mixing angle in M\o ller scattering},
\emph{Phys.\ Rev.\ Lett.}\ {\bf 95} (2005) 081601
[hep-ex/0504049].

\bibitem{Wang:2014bba} 
PVDIS Collaboration: D.~Wang \etal,
\emph{Measurement of parity violation in electron-quark scattering},
\emph{Nature} {\bf 506} (2014) 67.

\bibitem{Souder:2016xcn} 
P.~A.~Souder,
\emph{Parity violation in deep inelastic scattering with the SoLID spectrometer at JLab},
\emph{Int.\ J.\ Mod.\ Phys.\ Conf.\ Ser.}\ {\bf 40} (2016) 1660077.

\bibitem{Zeller:2001hh} 
NuTeV Collaboration: G.~P.~Zeller \etal,
\emph{A precise determination of electroweak parameters in neutrino nucleon scattering}, \\
\emph{Phys.\ Rev.\ Lett.}\ {\bf 88} (2002) 091802 
[Erratum {\em ibid.}\ {\bf 90} (2003) 239902]
[hep-ex/0110059].

\bibitem{Canas:2016vxp} 
B.~C.~Canas \etal,
\emph{The weak mixing angle from low energy neutrino measurements: a global update},
\emph{Phys.\ Lett.}\ {\bf B761} (2016) 450
[arXiv:1608.02671].

\bibitem{Wood:1997zq} 
C.~S.~Wood \etal,
\emph{Measurement of parity nonconservation and an anapole moment in cesium}, 
\emph{Science} {\bf 275} (1997) 1759.

\bibitem{Willmann:2012mxa} 
L.~Willmann \etal,
\emph{Trapped radioactive isotopes for fundamental symmetry investigations. The TRI$\mu$P Facility},
\emph{Hyperfine Interact.}\ {\bf 211} (2012) 39.
  
\bibitem{Kumar:2013yoa} 
K.~S.~Kumar, S.~Mantry, W.~J.~Marciano and P.~A.~Souder,
\emph{Low energy measurements of the weak mixing angle},
\emph{Ann.\ Rev.\ Nucl.\ Part.\ Sci.}\ {\bf 63} (2013) 237
[arXiv:1302.6263].

\bibitem{Erler:2013xha} 
J.~Erler and S.~Su,
\emph{The weak neutral current}, \\
\emph{Prog.\ Part.\ Nucl.\ Phys.}\ {\bf 71} (2013) 119
[arXiv:1303.5522].
  
\bibitem{Erler:2014fqa} 
J.~Erler, C.~J.~Horowitz, S.~Mantry and P.~A.~Souder,
\emph{Weak polarized electron scattering},
\emph{Ann.\ Rev.\ Nucl.\ Part.\ Sci.}\ {\bf 64} (2014) 269
[arXiv1401.6199].
  
\bibitem{Davier:2017zfy}
M.~Davier, A.~H\"ocker, B.~Malaescu and Z.~Zhang, 
\emph{Reevaluation of the hadronic vacuum polarisation contributions to the Standard Model predictions of 
the muon $g-2$ and $\alpha(M_Z)$ using newest hadronic cross-section data},
\emph{Eur.\ Phys.\ J.}\ {\bf C77} (2017) 827
[arXiv:1706.09436].

\bibitem{Jegerlehner:2011ti}
F.~Jegerlehner and R.~Szafron,
\emph{$\rho^0-\gamma$ mixing in the neutral channel pion form factor $F_\pi^e$ and its role in comparing 
$e^+e^-$ with $\tau$ spectral functions}, \\
\emph{Eur.\ Phys.\ J.}\ {\bf C71} (2011) 1632   
[arXiv:1101.2872].

\bibitem{Blum:2016xpd} 
RBC/UKQCD Collaboration: T.~Blum \etal,
\emph{Lattice calculation of the leading strange quark-connected contribution to the muon $g-2$},
\emph{JHEP} {\bf 1604} (2016) 063
[arXiv:1602.01767].
    
\bibitem{Chakraborty:2014mwa}
HPQCD Collaboration: B.~Chakraborty \etal,
\emph{Strange and charm quark contributions to the anomalous magnetic moment of the muon},
\emph{Phys.\ Rev.}\ {\bf D89} (2014) 114501
[arXiv:1403.1778].
      
\bibitem{Okubo:1963fa}
S.~Okubo, 
\emph{$\phi$-meson and unitary symmetry model}, 
\emph{ Phys.\ Lett.}\ {\bf 5} (1963) 165.

\bibitem{Zweig:1964jf}
G.~Zweig,
\emph{An $SU_3$ model for strong interaction symmetry and its breaking II}, \\
\emph{CERN preprint}, 8419/TH.412 (1964).

\bibitem{Iizuka:1966fk}
J.~Iizuka,
\emph{A systematics and phenomenology of meson family}, \\
\emph{Prog.\ Theor.\ Phys.\ Suppl.}\ {\bf 37} (1966) 21.
   
\bibitem{Blum:2015you} 
RBC/UKQCD Collaboration: T.~Blum \etal,
\emph{Calculation of the hadronic vacuum polarization disconnected contribution to the muon anomalous 
magnetic moment}, \\
\emph{Phys.\ Rev.\ Lett.}\ {\bf 116} (2016) 232002
[arXiv:1512.09054].
 
\bibitem{Erler:1998sy} 
J.~Erler,
\emph{Calculation of the QED coupling $\alpha(M_Z)$ in the modified minimal subtraction scheme},
\emph{Phys.\ Rev.}\ {\bf D59} (1999) 054008
[hep-ph/9803453].
  
\bibitem{Baikov:2012zm} 
P.~A.~Baikov, K.~G.~Chetyrkin, J.~H.~K\"uhn and J.~Rittinger,
\emph{Vector correlator in massless QCD at order $O(\alpha_s^4)$ and the QED beta-function at five loop},\\
\emph{JHEP} {\bf 1207} (2012) 017
[arXiv:1206.1284].

\bibitem{vanRitbergen:1997va}
T.~van~Ritbergen, J.~A.~M.~Vermaseren and S.~A.~Larin,
\emph{The four loop beta function in quantum chromodynamics},
\emph{Phys.\ Lett.}\ {\bf B400} (1997) 379
[hep-ph/9701390].

\bibitem{Czakon:2004bu}
M.~Czakon,
\emph{The Four-loop QCD beta-function and anomalous dimensions}, \\
\emph{Nucl.\ Phys.}\ {\bf B710} (2005) 485
[hep-ph/0411261].

\bibitem{Hall:1980kf} 
L.~J.~Hall,
\emph{Grand unification of effective gauge theories},
\emph{Nucl.\ Phys.}\ {\bf B178} (1981) 75.

\bibitem{Chetyrkin:1997un} 
K.~G.~Chetyrkin, B.~A.~Kniehl and M.~Steinhauser,
\emph{Decoupling relations to $O(\alpha_s^3)$ and their connection to low-energy theorems},
\emph{Nucl.\ Phys.}\ {\bf B510} (1998) 61
[hep-ph/9708255].

\bibitem{Chetyrkin:2006xg} 
K.~G.~Chetyrkin, J.~H.~K\"uhn and C.~Sturm,
\emph{Four-loop moments of the heavy quark vacuum polarization function in perturbative QCD},
\emph{Eur.\ Phys.\ J.}\ {\bf C48} (2006) 107
[hep-ph/0604234].
  
\bibitem{Kniehl:2006bf} 
B.~A.~Kniehl and A.~V.~Kotikov,
\emph{Heavy-quark QCD vacuum polarization function: Analytical results at four loops},
\emph{Phys.\ Lett.}\ {\bf B642} (2006) 68
[hep-ph/0607201].

\bibitem{Alemany:1997tn} 
R.~Alemany, M.~Davier and A.~H\"ocker,
\emph{Improved determination of the hadronic contribution to the muon (g-2) and to $\alpha(M_{Z}^2)$ using new data from hadronic $\tau$ decays}, \\
\emph{Eur.\ Phys.\ J.}\ {\bf C2} (1998) 123
[hep-ph/9703220].

\bibitem{Eidelman:1998vc} 
S.~Eidelman, F.~Jegerlehner, A.~L.~Kataev and O.~Veretin,
\emph{Testing nonperturbative strong interaction effects via the Adler function},
\emph{Phys.\ Lett.}\ {\bf B454} (1999) 369
[hep-ph/9812521].

\bibitem{Erler:2017dic} 
J.~Erler and R.~Ferro-Hern\'andez, 
\emph{Calculation of the QED coupling $\alpha(M_{Z}^2)$ in the $\msbar$ scheme: an update},
in preparation.
 
\bibitem{Patrignani:2016xqp} 
Particle Data Group: C.~Patrignani \etal,
\emph{Review of Particle Physics}, \\
\emph{Chin.\ Phys.}\ {\bf C40} (2016) 100001.

\bibitem{Erler:2000nx}
J.~Erler and M.~Luo,
\emph{Hadronic loop corrections to the muon anomalous magnetic moment},
\emph{Phys.\ Rev.\ Lett.}\ {\bf 87} (2001) 071804
[hep-ph/0101010].

\bibitem{Erler:2016atg} 
J.~Erler, P.~Masjuan and H.~Spiesberger,
\emph{Charm quark mass with calibrated uncertainty}, \\
\emph{Eur.\ Phys.\ J.}\ {\bf C77} (2017) 99
[arXiv:1610.08531].

\bibitem{Jegerlehner:2017zsb} 
F.~Jegerlehner,
\emph{Variations on Photon Vacuum Polarization}, [arXiv:1711.06089].

\bibitem{Hagiwara:2011af} 
K.~Hagiwara, R.~Liao, A.~D.~Martin, D.~Nomura and T.~Teubner,
\emph{$(g-2)_{\mu}$ and $\alpha(M_Z^2)$ re-evaluated using new precise data},
\emph{J.\ Phys.}\ {\bf G38} (2011) 085003 [arXiv:1105.3149].

\bibitem{Bernecker:2011gh}  
D.~Bernecker and H.~B.~Meyer,
\emph{Vector correlators in lattice QCD: methods and applications},
\emph{Eur.\ Phys.\ J.}\ {\bf A47} (2011) 148
[arXiv:1107.4388].

\bibitem{Keshavarzi:2018mgv} 
A.~Keshavarzi, D.~Nomura and T.~Teubner,
\emph{The muon $g-2$ and $\alpha(M_Z^2)$: a new data-based analysis},  
[arXiv:1802.02995].

\end{thebibliography}
\end{document}